\documentclass[preprint2]{aastex}

\shorttitle{X-ray Shell from 30~Dor~C}
\shortauthors{Bamba et al.}

\begin{document}

\title{Thermal and Non-thermal X-Rays from the LMC Super Bubble 30~Dor~C}
\author{
Aya Bamba, Masaru Ueno, Hiroshi Nakajima, and Katsuji Koyama,
}
\affil{
Department of Physics, Graduate School of Science, Kyoto University,
Sakyo-ku, Kyoto 606-8502, Japan}
\email{bamba@cr.scphys.kyoto-u.ac.jp,
masaru@cr.scphys.kyoto-u.ac.jp,
nakajima@cr.scphys.kyoto-u.ac.jp,
koyama@cr.scphys.kyoto-u.ac.jp}

\begin{abstract}

We report on the discovery of thermal and non-thermal X-rays
from the shells of the super bubble (SB)
30~Dor~C in the Large Magellanic Cloud (LMC).
The X-ray morphology is a nearly circular shell
with a radius of $\sim$40~pc,
which is bright on the northern and western sides.
The spectra of the shells are different from region to region.
The southern shell shows clear emission lines,
and is well fitted with a model of a thin-thermal plasma ($kT = 0.21$~keV)
in non-equilibrium ionization (NEI) plus a power-law component.
This thermal plasma is located inside of the H$\alpha$ emission,
which is the outer edge of the shell of the SB.
The northern and western sides of the SB are dim in H$\alpha$ emission,
but are bright in non-thermal (power-law) X-rays
with a photon index of 2.1--2.9.
The non-thermal X-ray shell traces
the outer boundary of the radio shell.  
These features of thin-thermal and non-thermal X-rays 
are similar to those of SN~1006,
a prototype of synchrotron X-ray shell,
but the non-thermal component of 30~Dor~C is about ten-times brighter
than that of SN~1006.
30~Dor~C is the first candidate of an extragalactic SB,
in which energetic electrons are accelerating in the shell.
The age is much older than that of SN~1006,
and hence the particle acceleration time
in this SB may be longer than those in normal shell-like SNRs.
We found point-like sources associated with some of tight star clusters.
The X-ray luminosity and spectrum are consistent with
those of young clusters of massive stars.
Point-like sources with non-thermal spectra are also found in the SB.
These may be background objects (AGNs)
or stellar remnants (neutron stars or black holes).

\end{abstract}


\section{Introduction}

Since the discovery of cosmic rays \citep{hess}, 
the sites and mechanisms of cosmic ray acceleration
up to the knee energy,
which is a spectral kink of cosmic rays at $\sim 10^{15}$~eV,
have been key problems.
The most plausible scenario is diffusive shock acceleration (DSA)
in the shock fronts of supernova remnants (SNRs)
\citep{bell,blandford};
in this scenario, particles receive energy
whenever they cross the shock from downstream to upstream.

\citet{koyama1995} discovered synchrotron X-rays
from the shock front of SN~1006,
and confirmed that
the SNR accelerates cosmic rays up to $\sim$TeV,
together with the discovery of inverse Compton TeV $\gamma$-rays
\citep{tanimori},
which helps us
to estimate the magnetic field and the maximum energy of electrons.
Non-thermal X-rays are the best tracer of cosmic ray acceleration sites.
In fact, several Galactic SNRs have already been confirmed as
cosmic ray accelerators using hard X-rays
(SN~1006; Koyama et al.\ 1995, Long et al.\ 2003, Bamba et al.\ 2003b,
G347.3$-$0.5; Koyama et al.\ 1997, Slane et al. 1999,
RCW~86; Bamba et al.\ 2000, Borkowski et al.\ 2001b, Rho et al.\ 2002,
G266.2$-$1.2; Slane et al.\ 2001,
Cas~A; Vink \& Laming\ 2003,
and Tycho; Hwang et al.\ 2002).
The {\it ASCA} Galactic plane survey has also found
some candidates \citep{bamba2001, ueno, bamba2003a}.
Although several Galactic SNRs have been confirmed
to be sites of cosmic ray acceleration,
some problems remain.
For example, no SNR has been found in which cosmic rays are accelerated 
up to the knee energy.
In fact, \citet{reynolds1999} have suggested that
the maximum energy of cosmic rays
accelerated in SNRs is at most $\sim 10^{14}$~eV, 
an order of magnitude below the knee energy.

Super bubbles (SBs) are large hot cavities
created by the combined actions of fast stellar winds
and successive supernova (SN) explosions of massive stars
in OB star associations.
Thus, SBs are strong candidates
for being sites where cosmic rays are accelerated to higher energies
(c.f., Bykov \& Fleishman\ 1992, Klepach, Ptuskin, \& Zirakashvili\ 2000).
However, it is difficult to study SBs in detail.
Their distances, and thus their sizes and energies,
remain uncertain in our Galaxy.
Furthermore, interstellar absorption often prevents observations
at optical, UV, and X-ray wavelengths.
The Large Magellanic Cloud (LMC) provides an ideal location
to examine the acceleration of cosmic rays in SBs
because of its proximity (50~kpc; Feast 1999),
low inclination angle (30\arcdeg--40\arcdeg; Westerlund 1997),
and low foreground absorption
($A_V < 0.3$~mag; Bessel 1991).

The LMC provides a sample of numerous SBs at a common distance
that is resolvable by modern X-ray detectors.
For this study, we singled out 30~Dor~C
as being the best target for our search.
A radio source was discovered by \citet{lemerne}
south-west of 30~Dor, and was named 30~Dor~C,
which is now categorized as a SB.
\citet{mills} found a shell-like structure in the 843~MHz band observation
with a radius of about 3\arcmin ($\sim$40~pc on 50~kpc distance).
Along the radio shell,
complex H$\alpha$, H$\beta$, and [SII] emissions were found,
which are bright in the south-east,
but dim in the other side \citep{mathewson}.
In the X-ray band,
{\it Einstein} detected the eastern shell of the 30~Dor region
for the first time \citep{long1981,chu}.
Dunne, Points, \& Chu (2001) (hereafter, DPC)
reported that the {\it ROSAT} spectrum 
required a thermal model with a rather high temperature ($\sim$1~keV).
\citet{itohM} found non-thermal X-rays in the {\it ASCA} data,
but could not spatially resolve the thermal and non-thermal components
in the SB.
Recently, \citet{dennerl} found a complete ring in soft and hard X-rays  
with a diameter of $\sim$6\arcmin with {\it XMM Newton}.
The hard X-rays from the shell resemble the synchrotron X-rays
in the shock front of SNRs, which is a site of cosmic ray acceleration.
Therefore, 30~Dor~C would be the first and good candidate of cosmic ray 
accelerating SBs.

In this paper,
we report on the first results of spatially resolved
hard X-ray spectral analyses of 30~Dor~C
using {\it Chandra} and {\it XMM-Newton} data.
In \S\ref{sec_obs}, we describe the data and their reduction.
We analyze the data in \S\ref{sec_results},
followed by a discussion in \S\ref{sec_discuss},
and a summary in \S\ref{sec_summary}.
The distance to the LMC is assumed to be 50~kpc \citep{feast}.

\section{Observations and Data Reduction}
\label{sec_obs}

{\it Chandra} observed SN~1987A several times
with the Advanced CCD Imaging Spectrometer (ACIS) S array;
the satellite has an excellent spatial resolution
($\sim$0.5\arcsec\ on the aim-point),
and one CCD has a field of view (FOV) of $\sim 8.5$\arcmin$\times$8.5\arcmin,
an energy range of 0.3--10.0~keV, and an energy resolution of $\sim$130~eV
at 5.9~keV.
The details for the satellite and the instruments are described
by \citet{garmire} and \citet{weisskopf}, respectively.
Some of the SN~1987A observations covered the 30~Dor~C region by chance,
which is located at about 5\arcmin\ northeast from  SN~1987A.
In order to study the diffuse structure of 30~Dor~C,
we selected two observations in which the grating instrument
is removed from the X-ray mirror axis
(Observation ID = 1044 and 1967; hereafter Obs.1 and 2).
The observed dates and targeted positions are given in Table~\ref{obslog}.
Data acquisition from the ACIS was made in the Timed-Exposure Faint mode
with a readout time of 3.24~s in both observations.
The data reductions and analyses were made using the {\it Chandra}
Interactive Analysis of Observations (CIAO) software version 2.3.
For a Level-1 processed events provided by a pipeline
processing at the {\it Chandra} X-ray Center,
we made a Charge Transfer Inefficiency (CTI) correction,
and selected {\it ASCA} grades 0, 2, 3, 4, and 6, as the X-ray events.
High-energy events due to charged particles and
hot and flickering pixels were removed.
There are some ``streaks'' in the CCD chip S4,
which are caused by a random deposition of a significant amount of charges
in pixels along the row during the serial read-out process.
These ``streak'' events were removed by using the program {\it destreak}%
\footnote{See http://asc.harvard.edu/ciao2.3/ahelp/destreak.html.}
in CIAO.
The total available times for each observation, after the screening,  
are listed in Table~\ref{obslog}.

{\it XMM-Newton} \citep{jansen}
also observed SN~1987A and the 30~Dor region several times;
the satellite has a spatial resolution of $\sim$14\arcsec\ 
and a relatively wide FOV with the radius of 15\arcmin.
We selected two observations
(Observation ID = 0104660301 and 0113020201; hereafter Obs.3 and 4),
which cover 30~Dor~C and are relatively free
from high background flares due to low-energy protons.
The observed dates and targeted positions are shown in Table~\ref{obslog}.
In both observations, only the metal oxide semiconductors (MOS) CCDs,
which have an energy range of 0.1--10.0~keV
and a similar energy resolution to {\it Chandra},
\citep{turner} were operated in the full-frame mode
with the medium filter \citep{stephan,villa} for blocking
ultra-violet photons.
The data reductions and analyses were made using
the {\it XMM-Newton} Standard Analysis System (SAS; Watson et al 2001)
version 5.4.1;
we performed the basic pipeline process following the SAS guide.
The background level was largely changed, particularly in Obs. 4,
hence we removed the data with a high background level
($>$0.6~cnts~s$^{-1}$ in the 10.0--15.0~keV band).
The exposure times in each observation after the screenings are listed
in Table~\ref{obslog}.

\section{Analyses and Results}
\label{sec_results}

\subsection{Diffuse Emission and Point-like Sources}
\label{sec_results_image}

\begin{figure}[hbtp]
\epsscale{1.0}
\plottwo{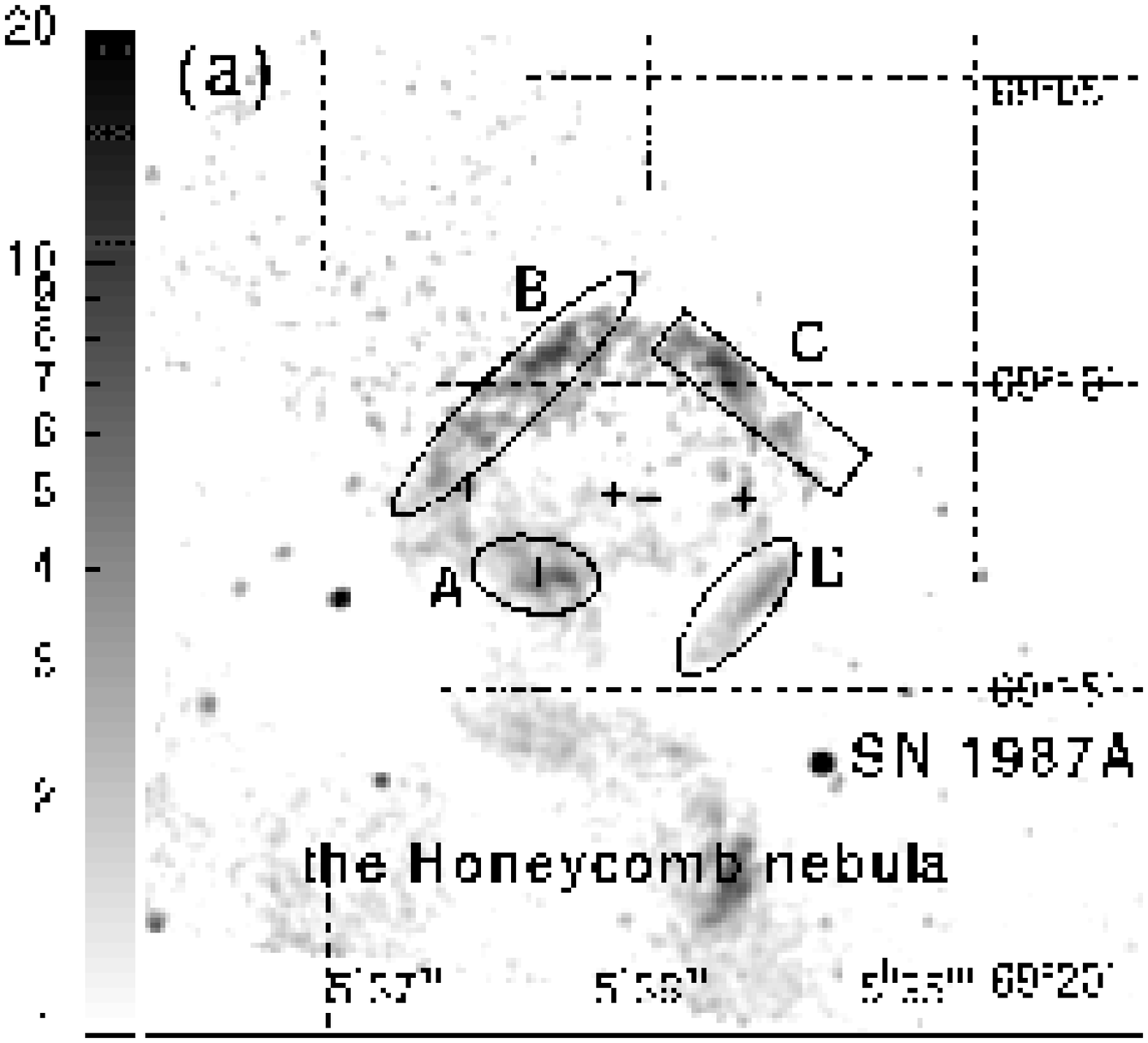}{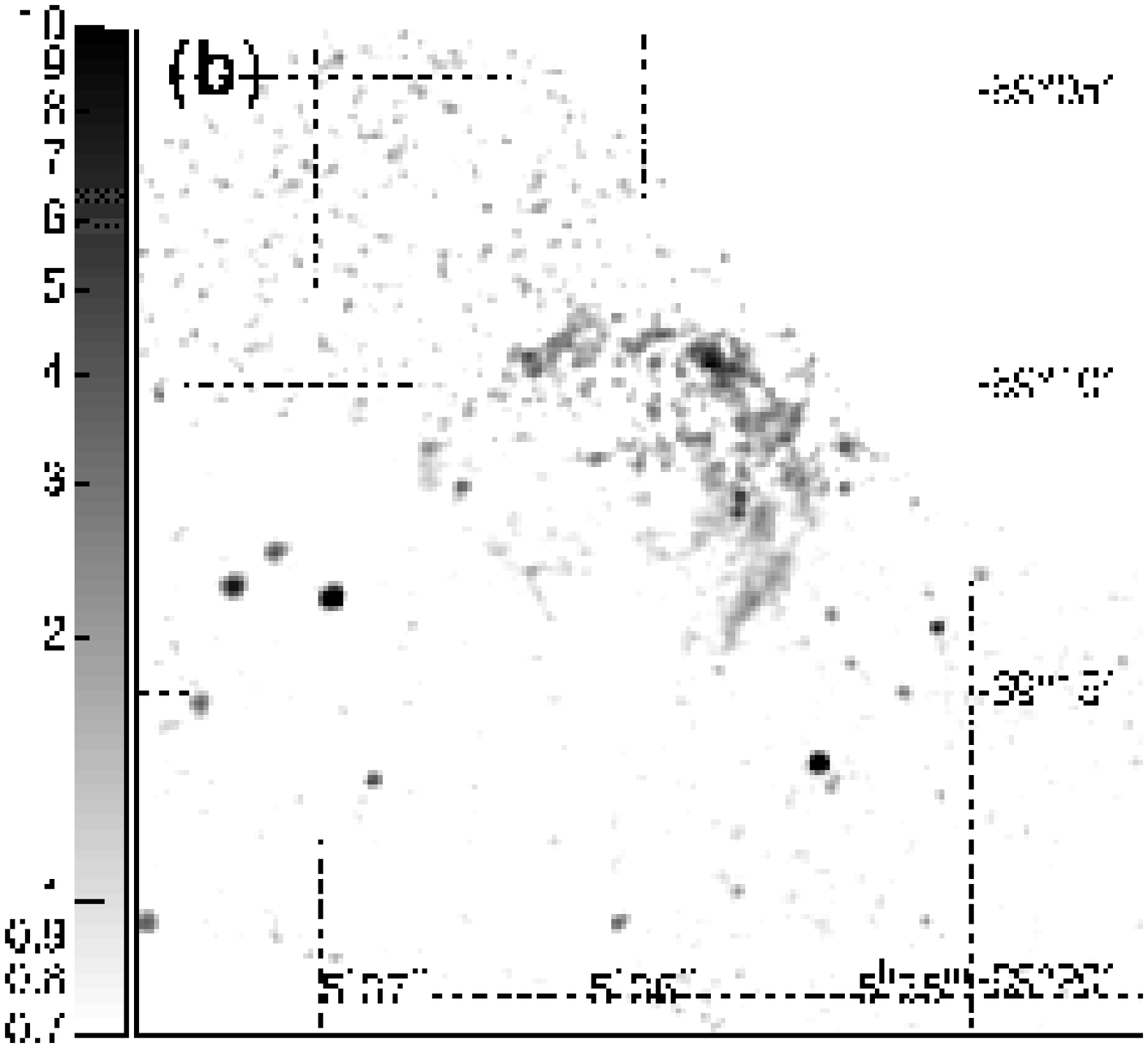}
\caption{{\it Chandra} ACIS images around 30~Dor~C
in the 0.7--2.0~keV band (a) and
in the 2.0--7.0~keV band (b),
with J2000 coordinates.
The scales are logarithmic
with the units of $\times 10^{-5}$~counts~s$^{-1}$arcmin$^{-2}$cm$^{-2}$,
as shown in the left bar for each image.
The source regions used for the spectral analyses are shown
by the solid lines in the left panel
(Fig.~\ref{shell_spec}, Table~\ref{spec_A} and \ref{spec_NT}).
The crosses represent the position of the detected point sources
shown in Table~\ref{point}.
The position of SN~1987A \citep{burrows,park,michael}
and the Honeycomb nebula \citep{dennerl}
are also shown.
\label{images}}
\end{figure}

Figure~\ref{images} shows
the soft (0.7--2.0~keV) and hard (2.0--7.0~keV) band {\it Chandra} images 
around 30~Dor~C, in which the two observations (Obs.1 and 2) are combined
with a correction of the exposure times.
A clear shell-like structure with a radius of $\sim$~170\arcsec\ 
($R\sim 40$~pc-radius at the 50~kpc distance) is seen in both bands.
In detail, however, the morphologies are different from each other;
the entire shell is seen in the soft band,
whereas the hard X-rays are visible only at the western part.
Catalogued SNRs, the Honeycomb nebula (SNR~0536$-$69.3) and SN~1987A
are also seen mainly in the soft band (see Figure~\ref{images}).
The X-ray features of these objects have been reported
with {\it XMM-Newton} \citep{dennerl}
and {\it Chandra} \citep{burrows,park,michael} monitoring observations.

In addition to the diffuse structure,
some point-like sources are found inside  30~Dor~C.
The exposure time of the  {\it XMM-Newton} observations are short, most of
the data suffer from a high background, and the spatial resolution
is not sufficient.
Hence, we concentrated on the {\it Chandra} data 
for the point-source search and analysis.  
At first, point sources were searched for with the {\it wavdetect} software%
in the 0.5--8.0~keV band images, then manually inspected
for any spurious point-like structure due mainly to a part of the
diffuse emission.
We thus found six point sources with a significance level of
$>$7.0$\sigma$, as shown in Figure~\ref{images} and Table~\ref{point}.
For these six point sources, we searched for optical, infrared, and radio 
counterparts,
and found that three (No.1, 3, and 4)
coincide at the positions of the brightest star clusters:
$\alpha$, $\beta$, and $\gamma$ \citep{lortet}.
We therefore checked further for any X-ray emission from the other clusters
($\delta$, $\epsilon$, and $\zeta$),
which are also members of the OB association LH~90 \citep{lucke}
encompassed by the 30~Dor~C shell.
However, we found no excess X-rays from these clusters at the 
3$\sigma$ limit.
We also searched for X-ray counterparts from {\it ROSAT} PSPC and HRI catalogues
\citep{haberl,sasaki}, but found no candidate.

\subsection{Spectra}

In the {\it Chandra} observations, 
we can see that the diffuse structure consists of several shell fragments
(see Figure~\ref{images}),
and the whole structure is widely spread over the two observed
regions with different configurations of the CCD types 
(back-illuminated and front-illuminated).
Therefore,
a spectral analysis on all of the diffuse structure is
technically and scientifically complicated.
For this reason, we divided the diffuse structure into four regions
(hereafter shells A--D),
as shown in Figure~\ref{images},
and performed the spectral analysis separately for each shell.
We excluded all of the detected point sources
(see \S\ref{sec_results_image}) from each shell.
The background regions were also selected separately for each shell
from the source-free regions in the same CCD and observations
as those of the shells.

\begin{figure}
\epsscale{0.45}
\plotone{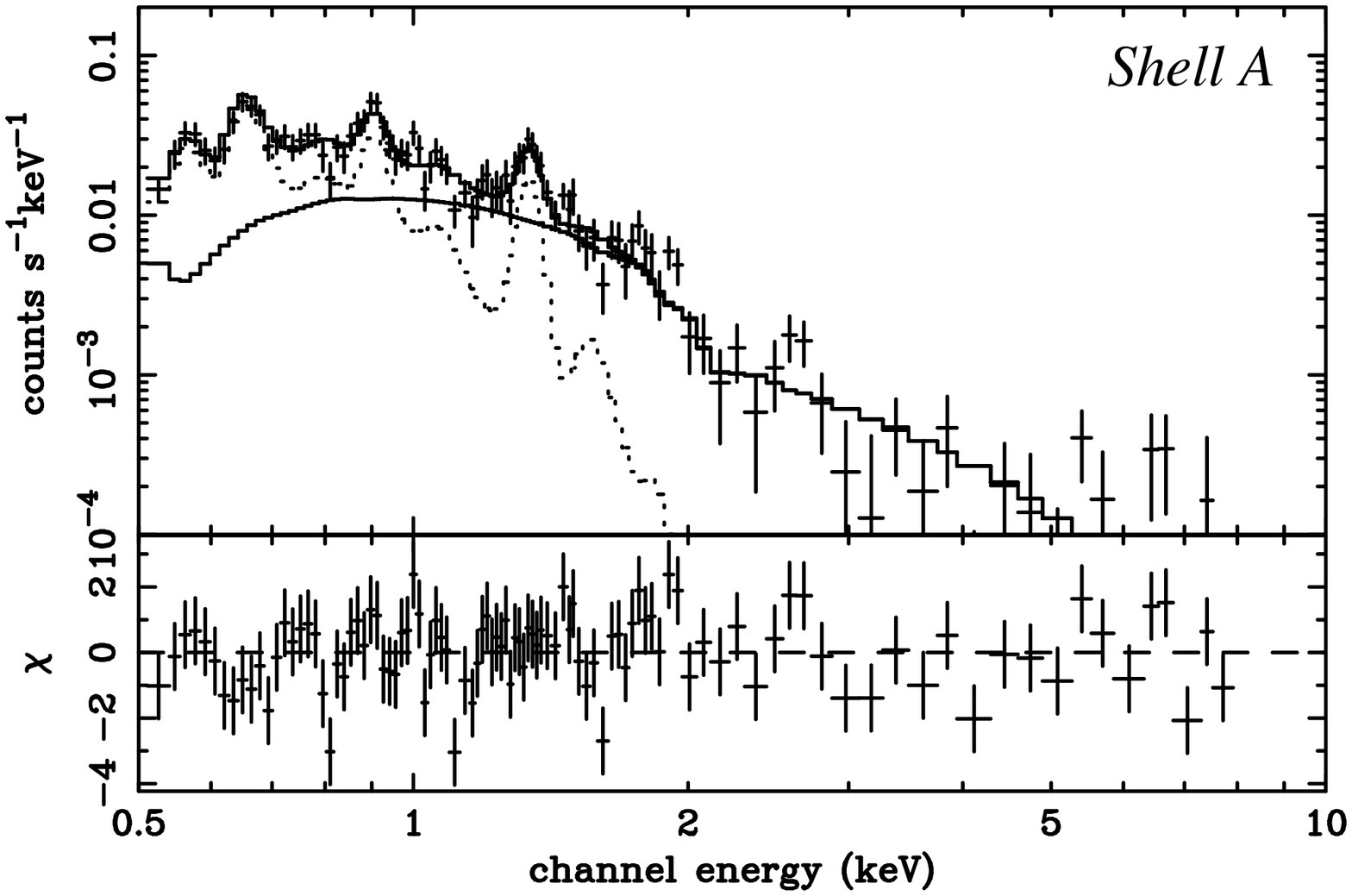}
\plotone{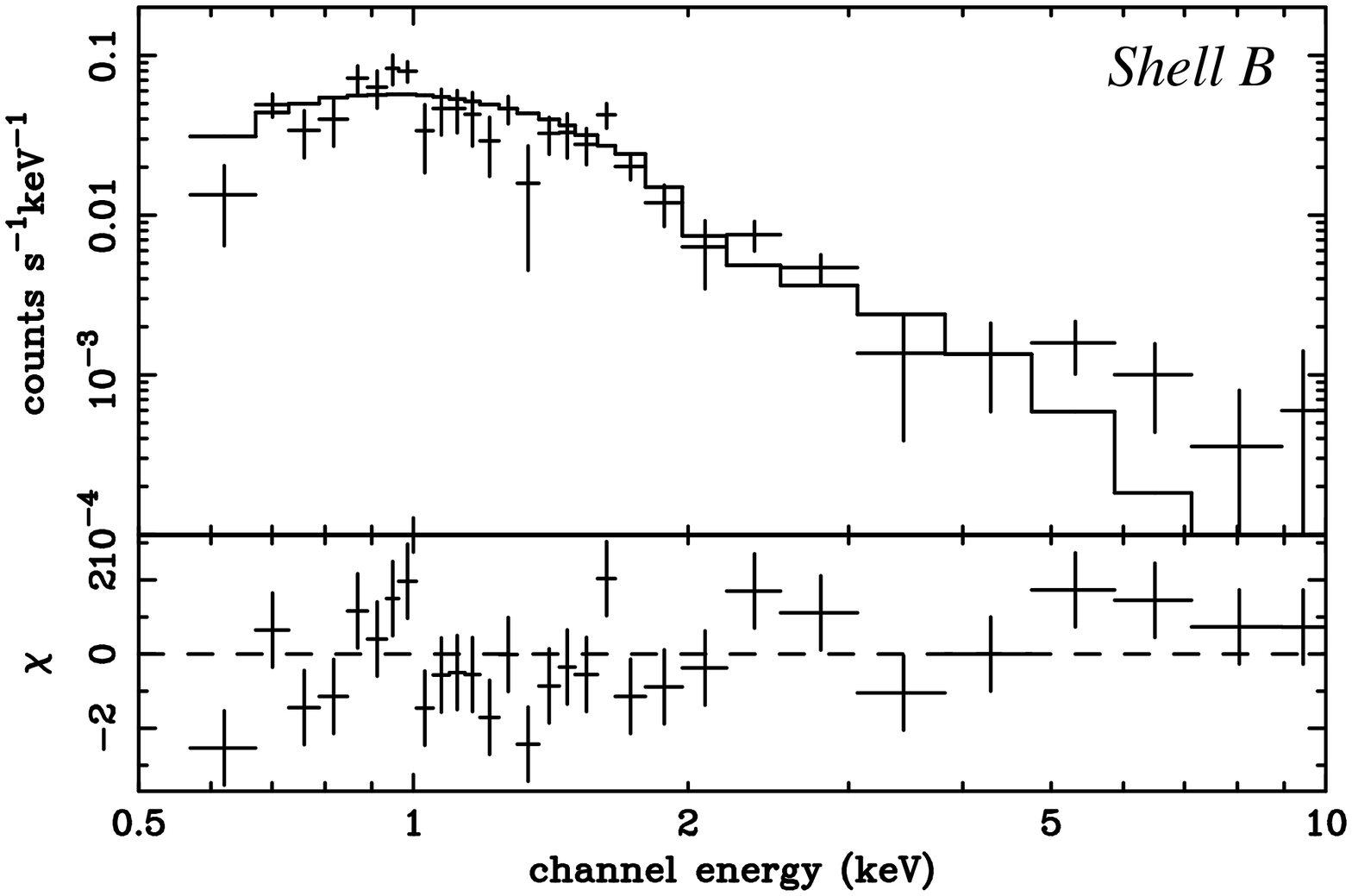}
\plotone{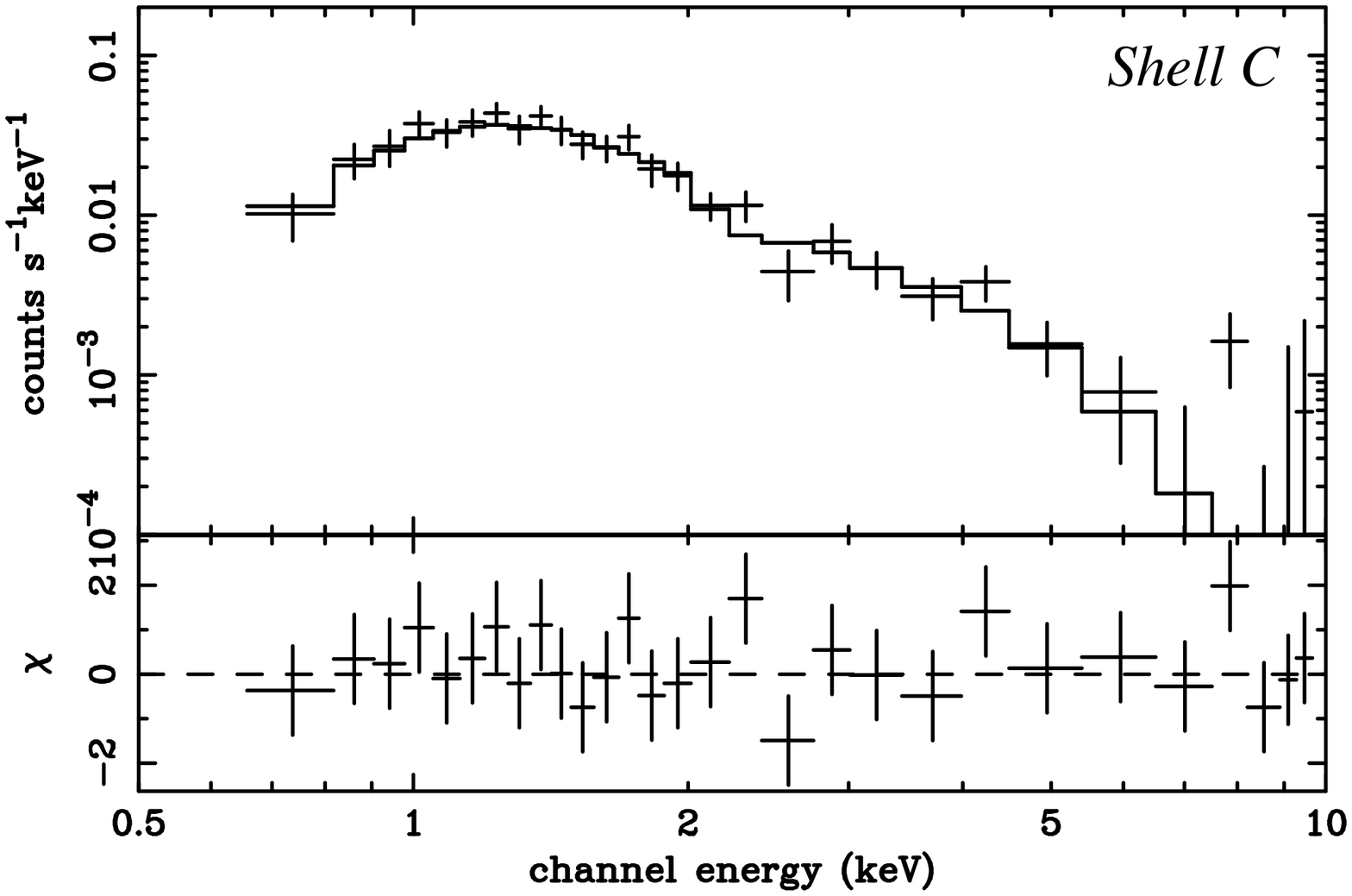}
\plotone{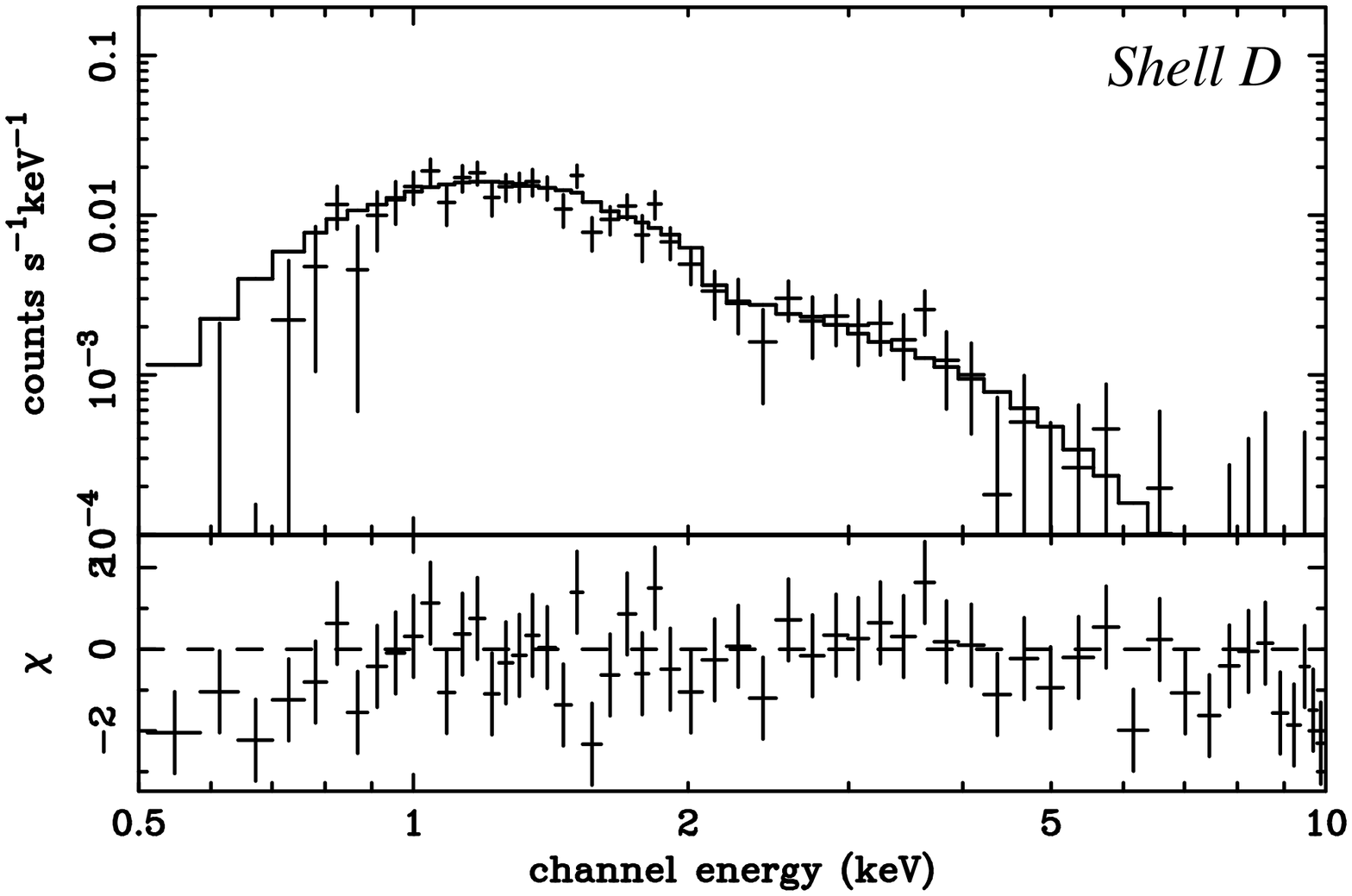}
\caption{Upper panels:
Background-subtracted spectra for the shells A--D (crosses).
Although the spectral fittings were  made
with both the {\it Chandra} and {XMM-Newton} data simultaneously,
only the {\it Chandra} data and results are shown for brevity.
The best-fit models are  shown
with solid (power-law) and dotted (NEI) lines.
Lower panels:
Data residuals from the best-fit models.
\label{shell_spec}}
\end{figure}

The {\it XMM-Newton} fields covered the whole diffuse structure.
However, for consistency with the {\it Chandra} analysis, we
divided the diffuse structure into shells~A--D, the same as
in the {\it Chandra} case.
The background regions were selected from the source-free regions
in the same observation.
The spectra of each shell taken from all the available
observations with {\it Chandra} and {\it XMM-Newton}
were analyzed simultaneously.
However, for brevity,
we show only the {\it Chandra} spectra (background-subtracted) and
the fitting results in Figure~\ref{shell_spec}.

For spectral analyses,
we used XSPEC v11.00
\citep{arnaud}
The spectrum of shell~A shows many line-like structures
with the center energies at 0.58, 0.68, 0.92, and 1.35~keV,
which correspond to the emission lines of He-like O K$\alpha$ and K$\beta$,
He-like Ne, and He-like Mg, respectively.
We therefore fit the spectrum with a thin-thermal plasma model in 
non-equilibrium ionization (NEI) ({\tt NEI}; Borkowski et al.\ 2001a)
with the mean LMC abundances \citep{russel,hughes}.
The interstellar absorption in our Galaxy and LMC
were treated separately.
The Galactic absorption column was estimated using the HI data by
\citet{dicky} as 
$N_{\rm H, HI} = 6.35\times 10^{20}$~cm$^{-2}$. 
\citet{arabadjis} reported that
the value of $N_{\rm H}$, measured in the X-ray band,
is twice that of $N_{\rm H, HI}$
in the case of $|b|>25^\circ$ and $N_{\rm H, HI}>5\times 10^{20}$~cm$^{-2}$.
Therefore, we fixed the galactic absorption column
to be $N_{\rm H} = 1.27\times 10^{21}$~cm$^{-2}$;
we used the cross sections by \citet{morrison}
and the solar abundances \citep{anders}.
The absorption column in the LMC was,
on the other hand,
treated as a free parameter with the mean LMC abundance \citep{russel,hughes}.

This thin-thermal plasma model was rejected
with a $\chi^2$/degree of freedom (d.o.f.)~=~331.8/211,
even if we allow the abundances to be free,
leaving a systematic data residual at the high-energy band.
We hence added a power-law component on the thin-thermal model.
Since the two-component model still leaves large residuals
at about 0.8~keV and 1.3~keV,
we allowed the abundance of Fe and Mg in the thermal plasma
(``NEI'' component) to be free.
The fitting was then greatly improved with a $\chi^2$/d.o.f = 264.6/209.
Although this two-component model is still rejected
in a statistical point of view,
further fine tuning on the model is beyond the scope of this paper.
Figure~\ref{shell_spec} and Table~\ref{spec_A} show
the best-fit models and parameters, respectively.

Unlike shell~A,
the X-ray spectra of shells~B--D are hard and featureless,
suggesting non-thermal origin.
In fact, a thin thermal model fitting requires
an unrealistically high temperature ($>$~2~keV)
and low abundances ($z < 0.3$). 
We therefore fitted the spectra with a power-law model
with absorption,
which was calculated in the same way as that for shell~A,
and found acceptable fits for all of the spectra.
The best-fit models and parameters are shown
in Figure~\ref{shell_spec} and Table~\ref{spec_NT}, respectively.

It is conceivable that the spectra of shells B--D may include
a small fraction of the thin-thermal component found in shell A.
We therefore added the same thin-thermal spectrum as that for shell~A,
and fitted with this composite model (thin-thermal plus power-law).
The free parameters are normalizations of the two components:
power-law index and $N_{\rm H}$ value.
However, no significant fraction of the thin thermal component is found
from shells~B--D.

For all of the point sources, the X-ray photons are collected
from an ellipse with the radii of the point spread function (PSF),
as listed in Table~\ref{point}.
We note that all of the sources are located far from the on-axis position of 
the X-ray mirror,
and hence the PSFs are larger than the best value of {\it Chandra} PSF
($\sim$ 0.5\arcsec on the aim point).
The background regions were selected from source-free regions
in the same way as the diffuse emissions.
We first fitted the spectra with a thin-thermal plasma model
in collisional equilibrium
({\tt MEKAL}; Mewe, Gronenschild, \& van den Oord\ 1985; Kaastra 1992)
with an absorption calculated in the same way as diffuse emission.
The abundances are fixed to be 0.3 solar,
the average value of interstellar medium in the LMC.
The fittings are acceptable for two sources (No.\ 1 and 4)
with reasonable temperature (2.1 and 1.0~keV),
but for the spectra of the other 4 sources,
the models are either rejected or
require an unreasonably high temperature.
We therefore fitted the spectra of these sources with a power-law model.
The best-fit parameters and the reduced $\chi^2$s are
listed in Table~\ref{point}.

\section{Discussion}
\label{sec_discuss}

\subsection{The absorption}

The absorption columns of the north-eastern shells (shells A and B)
are similar to those of most sources in the LMC ($\sim 10^{21}$~cm$^{-2}$),
whereas those of the other shells (C and D)
are significantly larger ($\sim 10^{22}$~cm$^{-2}$)
than the typical LMC absorption.
A similar trend was found for the point sources;
those in the western half, No.1 and 2, have a larger $N_{\rm H}$
($\sim 10^{22}$~cm$^{-2}$)
than those in the eastern sources, No.3--6 ($\leq 10^{21}$~cm$^{-2}$).
Since \citet{dunne} has already reported this tendency,
we have thus confirmed the results
with the better spatial and spectral 
capability of {\it Chandra}.
This systematic increase of absorption toward the western region of 30~Dor~C 
may be due to extra absorption of a molecular cloud 
located in front of the western half of 30~Dor~C.
To verify our conjecture,
we searched for the molecular cloud
in the CO map (Figure~2(a) in Yamaguchi et al.\ 2001)
and found the candidate with the intensity of $I$(CO) $\sim 3.6$~K~km~s$^{-1}$.
With a conversion factor of
$N({\rm H}_2)/I({\rm CO})
\sim 9\times 10^{20}$cm$^{-2}$(K~km~s$^{-1}$)$^{-1}$ \citep{fukui},
the estimated absorption column due to the molecular cloud is
$N_{\rm H}^{MC}\sim 6.5\times 10^{21}$~cm$^{-2}$,
which is consistent with our result.

\subsection{The Thermal Emission}

30~Dor~C is a SB made by a strong stellar wind
and/or successive supernova explosions of massive stars
located in the OB star association LH~90.
The age of this star association, or that of 30~Dor~C,
is on the order of a few to 10~Myr \citep{lucke}. 
The thermal emissions are enhanced in the south-eastern side of 30~Dor~C
(around shell~A).
The position of this component coincides with
that of the H$\alpha$ emission \citep{dunne}.
The plasma temperature of 30~Dor~C is rather high
compared to those of other LMC SBs \citep{dunne},
although it becomes significantly lower than previous results \citep{dunne}.
Perhaps due to the poor spectral resolution of {\it ROSAT},
previous observations could not resolve
the power-law component (hard spectrum) from the thermal emission.

The X-ray luminosity of 30~Dor~C is significantly lower than
that of the other SBs \citep{dunne}.
With the assumption that the plasma in shell~A distributes uniformly
in the ellipsoid with radii of
$58^{\prime\prime}\times 34^{\prime\prime}\times 34^{\prime\prime}$
(total volume $V = 1.2\times 10^{59}$~cm$^{3}$),
the mean density ($n_{\rm e}$), thermal energy ($E$)
and the age of the plasma ($t_{\rm p}$) were calculated
as follows
using the emission measure $E.M. = n_{\rm e}^2V$ and ionization time scale
(see Table~\ref{spec_A}):
\begin{eqnarray}
n_{\rm e} &=& 4.7\ ({\rm 4.2-4.8})\times 10^{-1}~[{\rm cm^{-3}}], \\
E &\simeq& 3n_{\rm e}kTV = 5.7\ ({\rm 4.6-6.4})\times 10^{49}~[{\rm ergs}], \\
t_{\rm p} &=& 6.7\ (>0.9)\times 10^5~[{\rm yrs}].
\end{eqnarray}
Together with the H$\alpha$ emission around shell~A \citep{dunne},
we infer that the shell of 30~Dor~C collides with dense matter
and temporally emits thermal X-rays with a relatively high temperature.
The overabundance of a light element (Mg) relative to heavier element (Fe)
may indicate that a type II SN occurred
\citep{tsujimoto}.
Thus, the progenitor is a massive star
which is a member of cluster LH~90 \citep{lucke} near the center of 30~Dor~C
(see section~\S\ref{sec:point}).

\subsection{The Non-Thermal Emission}

In the 843~MHz band, \citet{mills} found a clear radio shell,
which is brightest on the south-western side (around shells~C and D)
and dim on the eastern side.
The non-thermal X-ray emissions are enhanced at the radio bright shell.
This fact implies that the non-thermal X-rays are emitted
by the same mechanism as the radio band.
Therefore, the non-thermal X-rays are likely to be synchrotron radiation
from the accelerated electrons, like SN~1006 \citep{koyama1995}
and other SNRs,
which accelerate particles up to TeV.
The X-ray photon index of 2.1--2.9 (see Table~\ref{spec_NT}) is,
in fact, typical to synchrotron emissions.

To verify the synchrotron origin,
we fitted the X-ray spectra with a {\tt SRCUT} model,
which represents the synchrotron emission from 
electrons with an energy distribution of a power-law
plus exponential cut-off
\citep{reynolds1998,reynolds1999}.
Since the radio index data is not accurate
due to the large background and contamination of thermal emission
\citep{mathewson},
we fixed the spectral index ($\alpha$) at 1~GHz to be 0.5,
which is expected from the first-order Fermi acceleration
and similar to that of SN~1006 ($\alpha=0.57$; Allen, Petre, Gotthelf 2001).

The fittings were statistically acceptable,
and the best-fit parameters are listed in Table~\ref{spec_NT}.
The best-fit cut-off frequency is also similar to that of SN~1006,
although the age of 30~Dor~C may be on the same order as that of LH~90
(10~Myr; Lucke \& Hodge 1970),
which is far larger than SN~1006.
This implies that, depending on the environment,
the acceleration of high-energy electrons up to the knee energy
can continue for a far longer time than the previous consensus
for the SNR case ($\sim 10^3$~yrs; Reynolds and Keohane 1999).
The electron acceleration time may have been extended,
because successive supernova explosions in 30 Dor C,
possibly over the course of a few Myr, may more-or-less continuously 
produce high-energy electrons.
The total luminosity of the non-thermal component
($\sim 5.3\times 10^{35}$~ergs~s$^{-1}$) is
about 10-times larger than that of SN~1006 \citep{koyama1995}.
This large non-thermal flux in 30~Dor~C would be due to
the large energy supply by multiple successive supernova explosions.

The expected flux density at 1~GHz is
3.0 (2.4--4.6)$\times 10^{-2}$~Jy with $\alpha=0.5$,
which is significantly smaller than
the observed value (1.0~Jy; Mills et al.\ 1984).
This ``inconsistency'' is not relaxed,
even if we assume a larger radio index of $\alpha = 0.6$
(1.8 (1.4--2.5)$\times 10^{-1}$Jy, as shown in Table~\ref{spec_NT}).
We infer that the larger observed radio flux than that expected from X-rays
would be due to either the contamination of thermal radio flux
or a large uncertainty of the background level (see Mills et al.\ 1984). 

The H$\alpha$ emission \citep{dunne} is anti-correlated
with the non-thermal components.
Similar features have been observed
in some of other SNRs with synchrotron X-rays:
SN~1006 \citep{winkler} and RCW~86 \citep{smith}.
Because the H$\alpha$ region has higher density,
it may have a higher magnetic field.
Therefore, the maximum electron energy is
limited by the quick synchrotron energy loss,
leading to reduced non-thermal X-rays.  

\subsection{Point Sources}
\label{sec:point}

We have identified three point-like X-ray sources (No.~1, 3, and 4)  
to the tight clusters of massive stars $\alpha$, $\beta$, and $\gamma$,
respectively.
The spectrum of No.1 ($\alpha$) is fitted with
a thin thermal plasma model 
of $\sim$2.1~keV,
which is consistent with stellar X-rays from massive stars.  
The cluster $\alpha$ is the brightest of the three,
with X-ray luminosity of $\sim 10^{34}$~ergs~s$^{-1}$.
Optical spectroscopy of this cluster revealed that
it includes one red giant and one Wolf-Rayet (WR) star
(MG~41 and Brey~58; Lortet \& Testor 1984).
Hence, the latter is a possible counterpart of the 2~keV source.
The X-ray luminosity is near to the upper end of a massive star,
or its binary \citep{maeda}. 
Although the X-ray luminosity of No.3,
a counterpart of the star cluster $\beta$, is
also consistent with a massive young star, the best-fit photon index,
$\sim$2 (see Table~\ref{point}), 
is rather typical to a rotation powered neutron star,
probably a stellar remnant of an SN explosion in the active star cluster.
The spectrum of No.3 ($\gamma$) is soft
and well-fitted with a $kT = 1.0$~keV thermal plasma model,
similar to a cluster $\alpha$.
The star cluster $\gamma$ also includes one OB star
(Sk~$-69\fdg212$; Sanduleak 1970),
and the X-ray luminosity is consistent with 
that of a young massive star. 

The other three sources (No,2, 5, 6) have no counterpart
in optical, in the infrared band \citep{breysacher,lortet} nor
in the SIMBAD data base.
Their spectra are relatively hard,
and are consistent with being background active galactic nuclei (AGNs),
or stellar remnants (black hole or neutron star)
by successive SN explosions.

In order to constrain
the nature of the point-like X-ray sources with a power-law spectrum, 
we further examined the time variability with the Kolmogorov-Smirnov test
\citep{press}.
However, no significant time variability was found even between the 
two observations. 
For high resolution timing,
we examined the {\it Chandra} High Resolution Camera (HRC) data (ObsID = 738),
but no pulsation was found from any of these point sources.

\section{Summary}
\label{sec_summary}

(1) In the shell of the super bubble 30~Dor~C,
we resolved non-thermal and thermal X-rays
using {\it Chandra} and {\it XMM-Newton} data.

(2) The thermal emission concentrates on the south-eastern shell
of 30~Dor~C.
The spectrum is well represented by a thin thermal plasma model
with $kT = $0.21~(0.19--0.23)~keV
and $n_{\rm e} = $0.47~(0.42--0.48)~cm$^{-3}$.

(3) The non-thermal X-rays are located at
northern and western parts of the SB.
The power-law model is well-fitted to the spectra
with $\Gamma = $2.1--2.9, similar to that of the cosmic ray accelerating SNR,
SN~1006.
This is the first discovery of non-thermal X-rays
from the shells of SBs.
The total luminosity is ten-times larger than that of SN~1006.

(4) We found six point-like sources in 30~Dor~C.
Three sources are located in tight clusters of massive stars,
and therefore they may be active Wolf-Rayet stars or compact stars.
The other three have power-law spectra with no optical counterpart,
implying that they are background AGNs or compact remnants of SNe.

\acknowledgments

The authors thank to Drs.\ K.~Imanishi, M.~Itoh, Y.~Moriguchi, and T.~Yoshida
for their contribution to this study in an early phase.
Our particular thanks are due to the referee, Dr. Points,
for his fruitful comments.
This search made use of the SIMBAD database operated by CDS at Strasbourg,
France.
A.B. and M.U. are supported by JSPS Research Fellowship for Young Scientists.
This work is also supported by a Grant-in-Aid for the 21st Century COE
``center for Diversity and Universality in Physics''.

\onecolumn

\begin{deluxetable}{cccccc}
\tabletypesize{\scriptsize}
\tablecaption{Observation log around 30~Dor C.
\label{obslog}}
\tablewidth{0pt}
\tablecolumns{2}
\tablehead{
\colhead{} &
\colhead{Satellite} & \colhead{Obs ID} &
\colhead{Position (J2000)} & \colhead{Date} & \colhead{Exposure} \\
 & & & (RA, DEC) & (yyyy/mm/dd) & (ks)
}
\startdata
Obs.1 & {\it Chandra} & 1044 &
($05^{\rm h}32^{\rm m}22\fs0$, $-69^{\rm d}16^{\rm m}33\fs0$) &
2001/04/25 & 18\\
Obs.2 & {\it Chandra} & 1967 &
($05^{\rm h}32^{\rm m}22\fs0$, $-69^{\rm d}16^{\rm m}33\fs0$) &
2000/12/07 & 99\\
Obs.3 & {\it XMM-Newton} & 0104660301 &
($05^{\rm h}35^{\rm m}28\fs0$, $-69^{\rm d}16^{\rm m}11\fs0$) &
2000/11/25 & 21\\
Obs.4 & {\it XMM-Newton} & 0113020201 &
($05^{\rm h}37^{\rm m}47\fs6$, $-69^{\rm d}10^{\rm m}20\fs0$) &
2001/11/19 & 16\\
\enddata
\end{deluxetable}

\begin{deluxetable}{p{1.5pc}cccccccc}
\tabletypesize{\scriptsize}
\tablecaption{Point source data around 30~Dor C.\tablenotemark{a}
\label{point}}
\tablewidth{0pt}
\tablecolumns{2}
\tablehead{
\colhead{No.} & \colhead{Name} & \colhead{PSF\tablenotemark{b}} & \colhead{SM\tablenotemark{c}} & \colhead{$\Gamma/kT$} & \colhead{$N_{\rm H}^{\rm LMC}$\tablenotemark{d}} & \colhead{Flux\tablenotemark{e}} & \colhead{Luminosity\tablenotemark{f}} & \colhead{$\chi^2$/d.o.f.}\\
 & CXO~J05... & arcsec$\times$arcsec & & \nodata/[keV] & $[\times 10^{22}\ {\rm H cm^{-2}}]$ & [ergs cm$^{-2}$s$^{-1}$] & [ergs s$^{-1}$] }
\startdata
1\dotfill & 3542.4$-$691152 & $7.0\times 3.6$ & MK & 2.1 (1.4--3.6) & 3.5 (2.5--5.1) & $1.8\times10^{-14}$ & 1.2$\times 10^{34}$ & 9.53/12 \\
2\dotfill & 3542.9$-$691206 & $7.3\times 3.8$ & PL & 2.1 (1.4--3.2) & 7.7 (4.4--13) & $2.0\times 10^{-14}$ & 1.5$\times 10^{34}$ & 8.25/13 \\
3\dotfill & 3559.9$-$691150 & $7.1\times 4.5$ & PL & 1.7 (1.3--2.6) & 0.13 ($<$ 1.3) & $6.5\times 10^{-15}$ & 2.2$\times 10^{33}$ & 4.03/7 \\
4\dotfill & 3606.6$-$691147 & $8.3\times 6.6 $ & MK & 1.0 (0.19--1.3) & 0 ($<$ 1.9) & $2.8\times 10^{-15}$ & 1.1$\times 10^{33}$ & 8.64/9 \\
5\dotfill & 3620.7$-$691303 & $8.5\times 4.4$ & PL & 1.9 (1.5--2.4) & 0 ($<0.22$) & 8.9$\times 10^{-15}$ & 3.1$\times 10^{33}$ & 19.8/20 \\
6\dotfill & 3633.3$-$691140 & $10.7\times 8.5$ & PL & 1.8 (1.5--2.3) & 0.5 (0.08--1.1) & $2.0\times 10^{-14}$ & 7.8$\times 10^{33}$ & 21.4/18
\enddata
\tablenotetext{a}{Parentheses indicate single-parameter
90\% confidence regions.}
\tablenotetext{b}{The elliptic PSF size.}
\tablenotetext{c}{Spectral model; PL: power-law,
MK: thin thermal plasma in collisional equilibrium
\citep{mewe,kaastra} with the abundance of 0.3 solar.}
\tablenotetext{d}{Absorption column in the LMC. The abundance is assumed to the average LMC values \citep{russel,hughes}.}
\tablenotetext{e}{Absorbed flux in the 0.5--9.0~keV band.}
\tablenotetext{f}{Intrinsic luminosity (absorptions are removed) in the 0.5--9.0~keV band, at the LMC
distance of 50~kpc \citep{feast}.}
\tablecomments{%
No.1, 3 and  4  have the optical counterparts, the stellar 
clusters $\alpha$, $\beta$, and $\gamma$ respectively \citep{lortet}.
}
\end{deluxetable}

\begin{deluxetable}{p{8pc}cc}
\tabletypesize{\scriptsize}
\tablecaption{Best-fit parameters of the shell A.\tablenotemark{a}
\label{spec_A}}
\tablewidth{0pt}
\tablecolumns{2}
\tablehead{
\colhead{Parameters} & \colhead{NEI} & \colhead{Power-law}
}
\startdata
$kT$/$\Gamma$ [keV/---]\dotfill & 0.21 (0.19--0.23) & 2.9 (2.7--3.1) \\
$n_{\rm e}t_{\rm p}$\tablenotemark{b}\ [$10^{12}$cm$^{-3}$s]\dotfill & 9.9 ($>$1.4) & \nodata \\
$EM$\tablenotemark{c}\ [10$^{58}$~cm$^{-3}$]\dotfill & 2.6 (2.1--2.8) & \nodata \\
${\rm [Mg/H]}$\tablenotemark{d}\dotfill & 3.3 (2.5--4.0) & \nodata \\
${\rm [Fe/H]}$\tablenotemark{d}\dotfill & 0.10 (0.06--0.14) & \nodata \\
Flux\tablenotemark{e} [ergs~cm$^{-2}$s$^{-1}$]\dotfill & $6.8\times 10^{-14}$ & $8.4\times 10^{-14}$ \\
$N_{\rm H}^{\rm LMC}$\tablenotemark{f}\ [$\times 10^{22}$~cm$^{-2}$]\dotfill & 0.24 (0.21--0.28) & \nodata\tablenotemark{g}\\
\enddata
\tablenotetext{a}{Parentheses indicate single-parameter
90\% confidence regions.}
\tablenotetext{b}{Ionization time-scale, where
$n_{\rm e}$ and $t_{\rm p}$ are the electron density and age of the plasma.}
\tablenotetext{c}{Emission measure
$EM = \int n_{\rm e}n_{\rm p}dV \simeq n_{\rm e}^2V \sim n_{\rm e}^2V$,
where $n_{\rm p}$ and $V$ are
the proton density and the plasma volume,
respectively. The distance to LMC is assumed to be 50~kpc.}
\tablenotetext{d}{Abundance ratio relative to the solar value \citep{anders}.}
\tablenotetext{e}{Absorbed flux in the 0.5--9.0~keV band.}
\tablenotetext{f}{Absorption column in the LMC.
The abundances are assumed to be the average LMC values \citep{russel,hughes}.}
\tablenotetext{g} {Fixed to the same value as that for the NEI component.}
\end{deluxetable}

\begin{deluxetable}{p{12pc}ccc}
\tabletypesize{\scriptsize}
\tablecaption{Best-fit parameters of the shells B--D.\tablenotemark{a}
\label{spec_NT}}
\tablewidth{0pt}
\tablecolumns{2}
\tablehead{
 & \colhead{B} & \colhead{C} & \colhead{D}
}
\startdata
Power-law\\
\hspace*{3mm}Photon Index\dotfill & 2.7 (2.5--2.9) & 2.3 (2.1--2.4) & 2.5 (2.3--2.7) \\
\hspace*{3mm}Flux\tablenotemark{b}\ [ergs s$^{-1}$cm$^{-2}$]\dotfill & $4.7\times 10^{-13}$ & $4.5\times 10^{-13}$ & $1.5\times 10^{-13}$ \\
\hspace*{3mm}$N_{\rm H}^{\rm LMC}$ \ [10$^{22}$~cm$^{-2}$]\tablenotemark{c}\dotfill & 0.10 (0.06--0.19) & 1.1 (0.94--1.4) & 1.0 (0.85--1.3) \\
\hspace*{3mm}$\chi^2$/d.o.f.\dotfill & 193.0/138 & 68.3/84 & 121.5/110 \\
{\tt SRCUT} $(\alpha=0.5)$\\
\hspace*{3mm}$\nu_{\rm rolloff}$ [$10^{16}$~Hz]\dotfill & 6.2 (4.1--7.9) & 24 (15--47) & 12 (6.4--23) \\
\hspace*{3mm}Flux density at 1~GHz [$10^{-2}$Jy]\dotfill & 1.9 (1.5--3.0) & 0.71 (0.67--0.75) & 0.42 (0.25--0.83) \\
\hspace*{3mm}$N_{\rm H}^{\rm LMC}$\ [10$^{22}$~cm$^{-2}$]\tablenotemark{c}\dotfill & ($<$0.04) & 1.0 (0.90--1.1) & 0.89 (0.75--1.0) \\
\hspace*{3mm}$\chi^2$/d.o.f.\dotfill & 195.5/138 & 67.0/84 & 120.2/110\\
{\tt SRCUT} $(\alpha=0.6)$\\
\hspace*{3mm}$\nu_{\rm rolloff}$ [$10^{16}$~Hz]\dotfill & 8.0 (5.3--11) & 36 (20--75) & 16 (8.3--35) \\
\hspace*{3mm}Flux density at 1~GHz [$10^{-2}$Jy]\dotfill & 11 (8.2--16) & 4.1 (3.9--4.3) & 2.5 (1.6--4.6) \\
\hspace*{3mm}$N_{\rm H}^{\rm LMC}$\ [10$^{22}$~cm$^{-2}$]\tablenotemark{c}\dotfill & ($<$0.01) & 1.0 (0.91--1.2) & 1.1 (1.0--1.3) \\
\hspace*{3mm}$\chi^2$/d.o.f.\dotfill & 195.2/138 & 67.1/84 & 121.8/110
\enddata
\tablenotetext{a}{Parentheses indicate single-parameter
90\% confidence regions.}
\tablenotetext{b}{Flux in the 0.5--9.0~keV band.}
\tablenotetext{c}{Absorption column in the LMC.
The abundances are assumed to be the average LMC values
\citep{russel,hughes}.}
\end{deluxetable}

\end{document}